\documentclass[11pt]{article}

\usepackage{aas_macros,amsmath,amssymb,comment,cite,esint,graphicx,mathtools}
\usepackage[margin=.8in,letterpaper]{geometry}
\usepackage[colorlinks=true]{hyperref}
\usepackage[affil-it]{authblk}
\usepackage{subcaption}
\usepackage[utf8]{inputenc}
\usepackage{mathrsfs}
\usepackage{appendix}
\usepackage{amssymb}
\usepackage{float}
\usepackage{color}
\usepackage{cite}
\usepackage{hyperref}
\hypersetup{pageanchor=false}
\usepackage{indentfirst}
\usepackage{url}
\usepackage{float}
\usepackage{caption}
\usepackage[numbers,square,comma,sort&compress,merge]{natbib}
\usepackage{esint}
\usepackage{overpic}
\usepackage{graphicx}
\usepackage{epsf,amsmath,bbold,amsfonts,stmaryrd}
\usepackage{textcomp}
\usepackage{ulem}
\usepackage{tikz}

\numberwithin{equation}{section}
\let\originalleft\left
\let\originalright\right
\renewcommand{\left}{\mathopen{}\mathclose\bgroup\originalleft}
\renewcommand{\right}{\aftergroup\egroup\originalright}
\mathcode`\*="8000
{\catcode`\*=\active\gdef*{\mathclose{}\,\mathopen{}}}

\newcommand{\be}{\begin{equation}}
\newcommand{\ee}{\end{equation}}
\newcommand{\bea}{\setlength\arraycolsep{2pt} \begin{eqnarray}}
\newcommand{\eea}{\end{eqnarray}}
\newcommand{\nn}{\nonumber}

\newcommand{\eff}{{\text{eff}}}
\newcommand{\ma}{{\text{max}}}

\newcommand{\E}{{\mathcal{E}}}
\newcommand{\X}{{\mathcal{X}}}
\newcommand{\Y}{{\mathcal{Y}}}
\newcommand{\Z}{{\mathcal{Z}}}

\def\D{\Delta}

\def\nn{\nonumber}

\def\be{\begin{equation}}
\def\ee{\end{equation}}
\def\bag{\begin{aligned}}
\def\eag{\end{aligned}}
\def\bea{\begin{eqnarray}}
\def\eea{\end{eqnarray}}
\def\ba{\begin{array}}
\def\ea{\end{array}}

\def\bc{\begin{center}}
\def\ec{\end{center}}

\begin{document}
\title{Polarized images of charged particles in vortical motions around a magnetized Kerr black hole}

\author{
Zhenyu Zhang$^{1}$, Yehui Hou$^{1}$, Zezhou Hu$^{1}$, Minyong Guo$^{2\ast}$,
Bin Chen$^{1, 3, 4}$}
\date{}

\maketitle

\vspace{-10mm}

\begin{center}
{\it
$^1$Department of Physics, Peking University, No.5 Yiheyuan Rd, Beijing
100871, P.R. China\\\vspace{4mm}

$^2$ Department of Physics, Beijing Normal University,
Beijing 100875, P. R. China\\\vspace{4mm}

$^3$Center for High Energy Physics, Peking University,
No.5 Yiheyuan Rd, Beijing 100871, P. R. China\\\vspace{4mm}

$^4$ Collaborative Innovation Center of Quantum Matter,
No.5 Yiheyuan Rd, Beijing 100871, P. R. China\\\vspace{2mm}
}
\end{center}

\vspace{8mm}

\begin{abstract}
In this work, we study the  images of a Kerr black hole (BH) immersed in uniform magnetic fields, illuminated by the synchrotron radiation of charged particles in the jet. We particularly focus on the spontaneously vortical motions (SVMs) of charged particles in the jet region and investigate the polarized images of electromagnetic radiations from the trajectories along SVMs. We notice that there is a critical value $\omega_c$ for charged particle released at a given initial position and subjected an outward force, and once $|qB_0/m|=|\omega_B|>|\omega_c|$ charged particles can move along SVMs in the jet region. We obtain the polarized images of the electromagnetic radiations from the trajectories along SVMs. Our simplified model suggests that the SVM radiations can act as the light source to illuminate the BH and form a photon ring structure.
\end{abstract}

\vfill{\footnotesize $\ast$ Corresponding author: minyongguo@bnu.edu.cn}

\maketitle

\newpage
\baselineskip 18pt
\section{Introduction}\label{sec1}

The Event Horizon Telescope (EHT) has successfully captured the first horizon-scale images of the supermassive black holes (BHs) at the center of the elliptical galaxy M87 and the Galactic center in 2017 \cite{EventHorizonTelescope:2019dse, EventHorizonTelescope:2022wkp}. These observations significantly enhance our understanding of the BH physics at horizon-scales, which has accumulated increasing attention \cite{Gyulchev:2019tvk, Guerrero:2021ues, Hou:2022gge, Rosa:2022tfv, Qin:2020xzu, Rosa:2023hfm}. In particular, from the photon ring and the inner shadow \cite{Chael:2021rjo}, we are able to catch a glimpse of the physical environments around BHs. In addition, the polarized BH images released in 2021 give us more information about the emission profile and magnetic field at the horizon-scale \cite{EventHorizonTelescope:2021bee, EventHorizonTelescope:2021srq}.

In previous studies, many works focus on the emission from the accretion flows around BHs \cite{Gralla:2019xty,  Cunha:2019hzj, Hadar:2020fda,  Li:2021riw, Zhang:2021hit, He:2022yse,  Bisnovatyi-Kogan:2022ujt, Chen:2022qrw, Chakhchi:2022fls, Zeng:2021mok, Wang:2022yvi, Guo:2021bhr, Hu:2022lek, Chen:2022scf, Cardenas-Avendano:2022csp, Vincent:2022fwj, Peng:2020wun, Zhang:2023okw}. According to \cite{EventHorizonTelescope:2019pgp}, the EHT Collaboration also considered models that the radiation comes mainly from the jet around a BH and found some of those models has not been rule out by any other observations temporarily. Other works concentrating on the images of the jet in the vicinity of BHs can be found in \cite{Moscibrodzka:2015pda, Kawashima:2020rmr, Emami:2021ick, Broderick:2022tfu, Yang:2022jzi, Papoutsis:2022kzp}. Regarding the mechanisms for jet-launching, previous research has shown that energy can be extracted through the B-Z process \cite{Blandford:1977ds}, the B-P process \cite{Blandford:1982di} and others \cite{Begelman:1984mw, Singh:2016lmu}. In the B-Z process, the BH energy is extracted by the magnetic field threading the horizon, and transmitted along field lines in the form of Poynting flux, accelerating charged particles over distance \cite{Blandford:1977ds}. In the B-P process, energy is extracted from magnetized disks through magnetic field lines extended from the disk and into the distance \cite{Blandford:1982di}. Nevertheless, due to the lower density of jets compared to accretion disks, it remains uncertain whether the magnetohydrodynamics (MHD) description is effective in this context. Some studies suggest that jets composed of collisionless ionized particles can effectively explain relevant astronomical phenomena \cite{Ptitsyna:2015nta, Globus:2013bda, Lyutikov:2023crk}. 

Furthermore, Ruffini et al. proposed a novel theoretical framework known as the ``inner engine" \cite{Ruffini:2018mkp, Ruffini:2018aiq, Rueda:2022fgz}, which suggests that the gravitomagnetic interaction between a Kerr BH and an external magnetic field induces a large-scale electric field capable of accelerating charged particles to ultra-relativistic energies, thus generating a powerful jet flow. Similar mechanisms of charged particle acceleration have also been studied in\cite{Frolov:2010mi, Frolov:2011ea, Igata:2012js, Stuchlik:2015nlt, Tursunov:2018erf, Kolos:2020gdc, Chakraborty:2021bsb}. The inner engine with strong magnetic field is expected to explained radiations in the gamma-ray band \cite{Rueda:2022fgz}, as it can accelerate electrons to PeV energies, resulting in synchrotron radiation of high-energy photons. However, the potential to obtain highly resolved images offers an opportunity to uncover horizon-scale physics. Therefore, it is necessary to investigate observational signatures of the inner engine at the horizon-scale.

In this work, we reexamine the inner engine mechanism and present the condition for particle ejection, i.e., the condition to form spontaneously vortical motions (SVMs). Moreover, we investigate the horizon-scale imaging results for SVMs. Firstly, by employing the effective potential method \cite{misner1973gravitation, Kovar:2010ty, Kopacek:2010yr}, we plot the effective potentials that visually show how charged particles are accelerated and the configurations capable of ejecting particles. We introduce an effective force along the axis and identify sufficient conditions for forming the SVMs of charged particles in this mechanism. Next, we study the synchrotron radiations of the SVM particles and discuss its polarization. By using the numerical backward ray-tracing method, we further explore their polarized images received by a distant observer. In our simplified model, we show that there exists the ring structure produced by the multiple images of the SVMs.

The paper is organized as follows. In Sec. \ref{sec2}, we review the framework of the inner engine and study the vortical motions of charged particles in the jet region. In Sec. \ref{sec3}, we set up our model for imaging the SVM radiations and show the results. We close our work with a summary in Sec. \ref{sec4}. In this paper, we work in the geometrized unit with $G = c = 1$.

\section{SVMs of charged particles in the jet region}\label{sec2}

In this section, we focus on SVMs of charged particles in the spacetime containing a Kerr BH with a uniform magnetic field. In particular, we assume that the uniform magnetic field of interest is not strong, so the backreaction to the spacetime can be ignored. Precisely, we assume that the uniform magnetic field is described by the Wald solution, which satisfies the Einstein-Maxwell equations \cite{Wald:1974np}. In \cite{Ruffini:2018mkp, Ruffini:2018aiq, Rueda:2022fgz}, it has been shown that the system containing a Kerr BH and a Wald magnetic field can be provided as an inner engine to accelerate charged particles to ultra-relativistic energies along vortical motions in the jet region of the BH.

\subsection{Kerr BH and Wald electromagnetic field}
In this subsection, we would like to review the basic framework of the inner engine and present some important results appeared in \cite{Rueda:2022fgz}. We begin with the background spacetime, which is described by the Kerr metric. The line element takes the following form
\bea
\mathrm{d}s^2&=&-\left(1-\frac{2Mr}{\Sigma}\right)\mathrm{d}t^2+\frac{\Sigma}{\Delta}\mathrm{d}r^2+\Sigma\mathrm{d}\theta^2+\left(r^2+a^2+\frac{2Mra^2}{\Sigma}\mathrm{sin}^2\theta\right)\mathrm{d}\phi^2-\frac{4Mra}{\Sigma}\mathrm{sin}^2\theta\mathrm{d}t\mathrm{d}\phi\,,\nn\\
&=&g_{\mu\nu}\mathrm{d}x^\mu \mathrm{d}x^\nu
\eea
in the Boyer-Lindquist (BL) coordinates, and 
\bea
\Delta=r^2-2Mr+a^2\,,\quad \Sigma=r^2+a^2\cos^2\theta\,,
\eea
where $M$ and $a$ are the mass and the spin parameters of the Kerr BH, respectively. For simplicity and without loss of generality, we set $M=1$ hereafter. The horizons can be found by solving the equation $\Delta=0$ which gives $r_\pm=1\pm\sqrt{1-a^2}$. 

Note that the Kerr BH is stationary and axisymmetric, we may consider a stationary uniform magnetic field, which was first found by R. Wald in \cite{Wald:1974np}. In the Wald solution, the electromagnetic 4-potential is given by 
\bea\label{at}
A_t&=&-aB_0\left[1-\frac{r}{\Sigma}\left(1+\cos^2\theta\right)\right]\,,\nn\\
A_\phi&=&\frac{1}{2}B_0\sin^2\theta\left[r^2+a^2-\frac{2ra^2}{\Sigma}\left(1+\cos^2\theta\right)\right]\,,
\eea
where the parameter $B_0$ denotes the magnetic strength at infinity. We disregard the backreaction of the magnetic field to the background spacetime. As shown in the work \cite{Rueda:2022fgz}, due to the interaction of the magnetic field and the gravitational field, an electric field is induced in the locally non-rotating frame (LNRF). The electric field is sourced by a quadrupolar distribution of surface charge density on the horizon \cite{1988Black}
\bea
\sigma(\theta)=\frac{B_0ar_+(r_+-1)\left(r_+\sin^4\theta-\cos^4\theta-\cos^2\theta\right)}{4\pi\left(r_+^2+a^2\cos^2\theta\right)}\,,
\eea
leading to a zero net charge of the BH. The tetrad basis of the LNRF takes the form
\bea\label{tetrad}
e_{(0)}=\frac{g_{\phi\phi}\partial_t-g_{\phi t}\partial_{\phi}}{\sqrt{g_{\phi\phi}\left(g_{\phi t}^2-g_{\phi\phi}g_{tt}\right)}}\,,\quad e_{(1)}=\frac{\partial_r}{\sqrt{g_{rr}}}\,,\quad e_{(2)}=\frac{\partial_\theta}{\sqrt{g_{\theta\theta}}}\,,\quad e_{(3)}=\frac{\partial_\phi}{\sqrt{g_{\phi\phi}}}\,.
\eea
The components of the electric and magnetic field measured by the locally non-rotating observer (LNRO) $u^\mu\equiv e^\mu_{(0)}$ in the LNRF are then given by 
\bea
E_{(i)}=F_{\mu\nu}e^\mu_{(i)}u^\nu=F_{\mu\nu}e^\mu_{(i)}e^\nu_{(0)}\,,\quad B_{(i)}=\frac{1}{2}\epsilon_{\mu\nu\sigma\rho}F^{\sigma\rho}u^\mu e^\nu_{(i)}=\frac{1}{2}\epsilon_{\mu\nu\sigma\rho}F^{\sigma\rho}e^\mu_{(0)}e^\nu_{(i)}\,,
\eea
where $F_{\mu\nu}=\partial_\mu A_\nu-\partial_\nu A_\mu$ is the electromagnetic field tensor in the BL coordinates and $\epsilon_{\mu\nu\sigma\rho}$ denotes the Levi-Civita tensor. After some monotonous calculations, the expressions of the non-zero components of the electric and magnetic field in the LNRF are 
\bea
E_{(1)}&=&-\frac{B_0a}{\Sigma^{2}A^{1/2}}\left[\left(r^2+a^2\right)\left(r^2-a^2\cos^2\theta\right)\left(1+\cos^2\theta\right)-2r^2\sin^2\theta\Sigma\right]\,,\nn\\
E_{(2)}&=&\frac{B_0a\D^{1/2}}{\Sigma^{2}A^{1/2}}2ra^2\sin\theta\cos\theta\left(1+\cos^2\theta\right)\,,\nn\\
B_{(1)}&=&-\frac{B_0\cos\theta}{\Sigma^2A^{1/2}}\left[2ra^2\left(2r^2\cos^2\theta+a^2+a^2\cos^4\theta\right)-\left(r^2+a^2\right)\Sigma^2\right]\,,\nn\\
B_{(2)}&=&-\frac{\Delta^{1/2}B_0\sin\theta}{\Sigma^2A^{1/2}}\left[a^2\left(r^2-a^2\cos^2\theta\right)\left(1+\cos^2\theta\right)+r\Sigma^2\right]\,,
\eea
where $A=(r^2+a^2)^2-\D a^2 \sin^2\theta$. Combining with Eq. (\ref{tetrad}), in the BL coordinate basis the electric and magnetic field are expressed as
\bea
E^r=\frac{\Delta^{1/2}}{\Sigma^{1/2}}E_{(1)}\,,\quad E^\theta=\frac{E_{(2)}}{\Sigma^{1/2}}\,,\quad B^r=\frac{\Delta^{1/2}}{\Sigma^{1/2}}B_{(1)}\,,\quad B^\theta=\frac{B_{(2)}}{\Sigma^{1/2}}\,.
\eea

\begin{figure}[h!]
  \centering
  \includegraphics[width=6.2in]{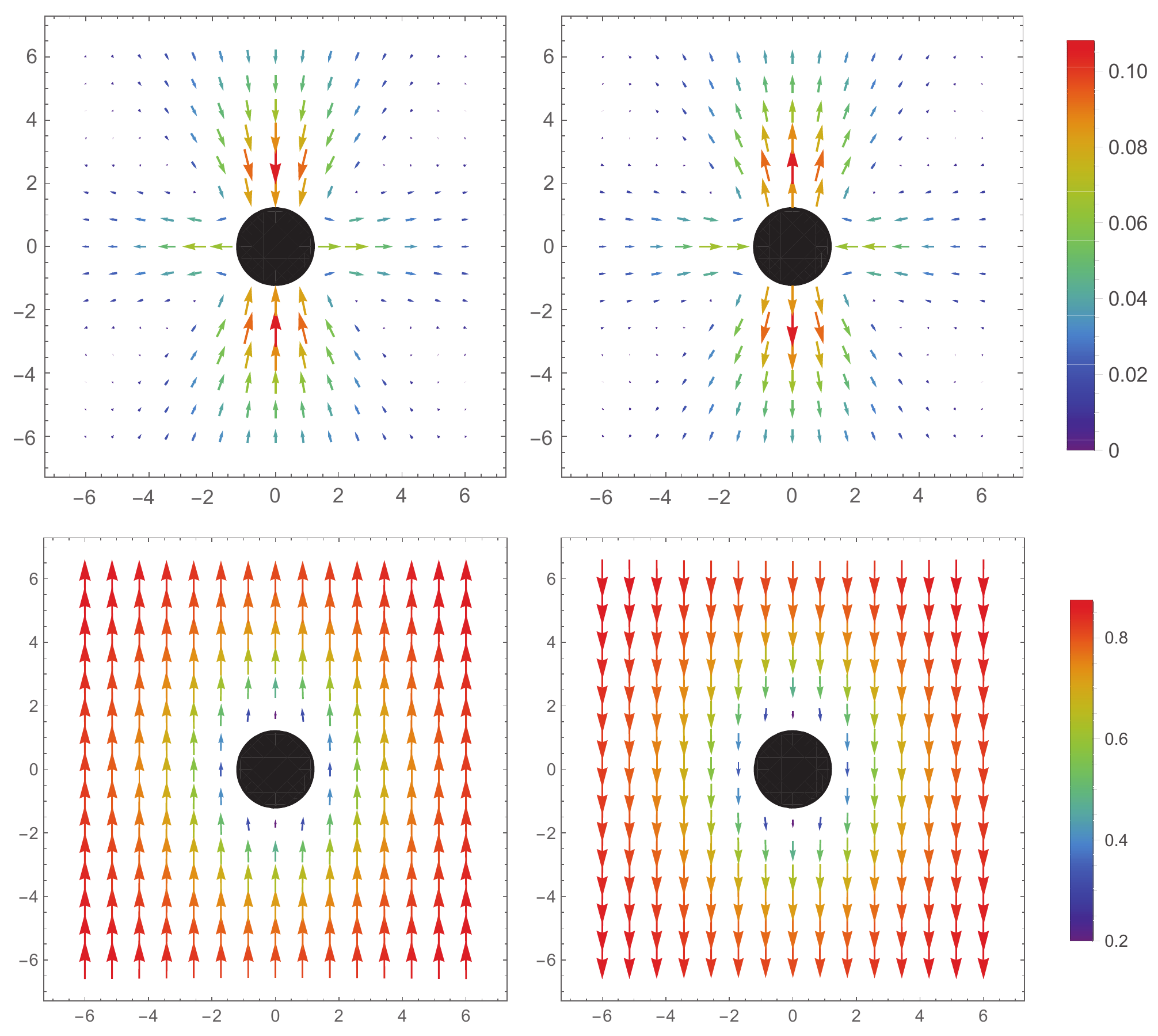}
  \centering
  \caption{Electromagnetic field configuration. Upper: Electric field lines of the Wald electromagnetic field in the $\rho$-$z$ plane with $\rho=r\sin\theta, z=r\cos\theta$. Lower: Magnetic field lines in the $\rho$-$z$ plane. Left: We set $a=0.94, B_0=1$. Right: $a=0.94, B_0=-1$. The black filled disk denotes the BH horizon.}
  \label{EBs}
\end{figure}

In Fig. \ref{EBs}, we show the electric and magnetic fields measured by the LNRO in the $\rho$-$z$ plane, where $\rho=r\sin\theta, z=r\cos\theta$. Our results are consistent with Fig.3 and Fig.4 in \cite{Rueda:2022fgz}, where their electricmagnetic field lines are presented in the Kerr-Schild coordinates\footnote{In the Kerr-Schild coordinates, the spatial coordinates $(x, y, z)$ are related to the BL ones by $x=\left(r\cos\phi-a\sin\phi\right)\sin\theta, y=\left(r\sin\phi+a\cos\phi\right)\sin\theta, z=r\cos\theta$. In \cite{Rueda:2022fgz}, the electromagnetic field lines in Figs. 3 and 4 are shown in the $x$-$z$ plane. In particular, when $\phi=0$, one has $x=r\sin\theta=\rho$, that is, our results coincide with theirs on the $\phi=0$ plane.}. Considering the $\mathcal{Z}_2$ symmetry in the Kerr BH spacetime, we discuss the behaviors of the electricmagnetic field in terms of the northern hemisphere. On the one hand, for the electric field we can see that there exists a critical angle $\theta_c$, at which the electric field vanishes and reverse direction in the upper panel of Fig. \ref{EBs}. In fact, we can find the value of $\theta_c$ from
\bea
\sigma(\theta_c)=0\,.
\eea
Note that the electric field force on electrons is opposite to the direction of the electric field, and it is just the other way around for protons. Therefore, when $B_0>0$ we have a polar electronic jet region within the critical angle, that is, $0\le\theta\le\theta_c$. On the contrary, a polar protonic jet region happens at $0\le\theta\le\theta_c$. when $B_0<0$.  On the other hand, in the lower panel of Fig. \ref{EBs} we can see that the magnetic field is asymptotically directed along the $z$-axis, since $|B^z|\gg |B^x|$ always holds. 

As a result, within the jet region, one expects that charged particles could be accelerated outward to high energies by the electric field and move vortically under the influence of the magnetic field in the jet region. If a charged particle is released without radial and polar velocities in the LNRF  and then move along an outward vortical motion in the jet region, we refer to the motion of the particle as a spontaneously vortical motion (SVM). The conical surface $\theta=\theta_c$ serves as the boundary of the SVM cluster.

\subsection{SVMs in the jet region}\label{2sub2}
In this subsection, we study SVMs of charged particles in the jet region and present the corresponding trajectories. We consider a charged particle of mass $m$ and charge $q$ in the Kerr BH spacetime with a Wald magnetic field. The Hamiltonian for the dynamics of the charged particle can be written in the form
\bea
H=\frac{1}{2}g_{\mu\nu}\dot{x}^\mu\dot{x}^\nu=\frac{1}{2}g^{\mu\nu}\left(\frac{\pi_\mu}{m}-\frac{q}{m}A_\mu\right)\left(\frac{\pi_\nu}{m}-\frac{q}{m}A_\nu\right)\,,
\eea
where $\dot{x}^\mu\equiv dx^\mu/d\tau$ with $\tau$ the proper time and $\pi_\mu=p_\mu+q A_\mu=m\dot{x}_\mu+qA_\mu$ is the canonical momentum with $p^\mu$ the 4-momentum. Then the motions of charged particles are governed by the Hamilton canonical equation, that is 
\bea\label{eom}
\dot{x}^\mu=\frac{\partial H}{\partial\pi_\mu}\,,\quad \dot{\pi}_\mu=-\frac{\partial H}{\partial x^\mu}\,.
\eea
Due to the presence of the Killing vectors $\partial_t$ and $\partial_\phi$ in the system, we can define the generalized energy and angular momentum as
\bea\label{EL}
E=-\frac{\pi_t}{m}=-\frac{p_t+qA_t}{m}\,,\quad L=\frac{\pi_\phi}{m}=\frac{p_\phi+qA_\phi}{m}\,,
\eea
which are constants along the motions of charged particles. In addition, in some of the discussions  bellow, we have scaled these two conserved quantities divided by the mass, that is, $E$ and $L$ are understood as the the generalized energy and momentum per unit mass. Incorporated with the normalization condition $g_{\mu\nu}\dot{x}^\mu\dot{x}^\nu=-1$, Eq.~\eqref{EL} implies 
\bea
E=\frac{-\beta+\sqrt{\beta^2-4\alpha X}}{2\alpha}\,,
\eea
where we introduce
\bea
\alpha&=&-g^{tt}\,,\nn\\
\beta&=&2\left[g^{t\phi}\left(L-\frac{q}{m}A_\phi\right)-g^{tt}\frac{q}{m}A_t\right]\,,\nn\\
X&=&-g^{\phi\phi}\left(L-\frac{q}{m}A_\phi\right)^2-g^{tt}\frac{q^2}{m^2}A_t^2+2g^{t\phi}\frac{q}{m}A_t\left(L-\frac{q}{m}A_\phi\right)-1-g^{\theta\theta}\frac{p_\theta^2}{m^2}-g^{rr}\frac{p_r^2}{m^2}\,.
\eea

Considering that we focus on the SVMs in the jet region, we should set the initial radial and polar velocities of the particle vanishing in the LNRF, that is, $p_{(1)}=p_{(2)}=0$ initially, or equivalently, $p_r(\tau=0)=p_\theta(\tau=0)=0$ \footnote{An analytical study on the vortical motions in the near horizon region of an extreme Kerr BH can be found in a recent work \cite{Hou:2023hto}.}. Hence, it is helpful to introduce an effective potential to analysis the SVMs of charged particles, which is defined as
\bea
V_{\eff}=E\vert _{p_\theta=p_r=0}=\frac{-\beta+\sqrt{\beta^2-4\alpha  \gamma}}{2\alpha} \, ,
\label{Veff}
\eea
where
\be
\gamma\equiv X\vert _{p_\theta=p_r=0}=-g^{\phi\phi}\left(L-\frac{q}{m}A_\phi\right)^2-g^{tt}\frac{q^2}{m^2}A_t^2+2g^{t\phi}\frac{q}{m}A_t\left(L-\frac{q}{m}A_\phi\right)-1\,.
\ee
Then, with the expression of the Wald solution Eq. (\ref{at}), we find that $V_{\eff}$ depends on the spin $a$, the generalized momentum $L$, and the parameter $\omega_B\equiv qB_0/m$, where $\omega_B$ is introduced as the coupling factor describing the interaction between charged particles and the electromagnetic field. In this work, we focus on $\omega_B<0$, namely, the charge and the magnetic field have the opposite sign. From Fig. \ref{EBs}, we can see that charged particles always have an outward electric field force in the jet region at $\omega_B<0$ so that the SVMs become possible when the electric field force is large enough to break through the gravitational barrier no matter whether the charge is positive or negative. In order to see the magnitudes of $\omega_B$ for electrons and protons, we recover the dimension of $\omega_B$ in Gaussian units, that is,
\bea\label{wBB}
w_B=\frac{\vert e\vert }{m_e}B_0\left(\frac{q}{\vert e\vert }\right)\left(\frac{m_e}{m}\right)\frac{GM}{c^3}\simeq86\left(\frac{q}{\vert e\vert }\right)\left(\frac{m_e}{m}\right)\left(\frac{M}{M_{\odot}}\right)\left(\frac{B_0}{1\text{Gauss}}\right)\,,
\eea
where $|e|$ is the unit charge, $m_e$ is the mass of an electron, $M_\odot$ is the solar mass. For the supermassive black hole in the centre of the M87 galaxy, the mass is about $6.5\times10^9M_\odot$ and the magnetic field strength is about $1-30$ Gauss according to the estimation by the EHT Collaboration \cite{EventHorizonTelescope:2021srq}. So let us take $B_0=10$ Gauss in our analysis. Thus, for an electron we have $|\omega_B|\simeq5.59\times10^{12}\gg1$ considering $|e|/m_e\simeq1.76\times10^{11}$ C/kg, as a result, electromagnetic field interactions are much larger than gravitational interactions. In addition, we  are also interested in the case of $|\omega_B|\sim1$ in terms of theoretical research, since the electromagnetic force is about as strong as the gravitational force in this case. Now the specific charge could be $|q|/m\sim10^{-2}$ C/kg,  which corresponds to charged dust or cloud of charged particles around the BH. In particular, considering that the electromagnetic field interactions vary with the specific charge and the gravitational force on a charged particle depends only on the mass, there might be a critical value $\omega_B=\omega_c$, at which the gravitational and electromagnetic forces just cancel out along the $z$ direction when we release the charged particle. In order to determine $\omega_c$, we define an effective force as 
\bea
F_z^\eff\equiv-\partial_zV_\eff\,,
\eea
along the $z$ direction. Thus, $\omega_c$ can be obtained by solving $F_z^\eff=0$ at the initial position $(t_i=0, r_i, \theta_i, \phi_i)$. Note that the system of interest is axisymmetric and we can set $\phi_i=0$ for simplicity. 

\begin{figure}[h!]
  \centering
  \includegraphics[width=6in]{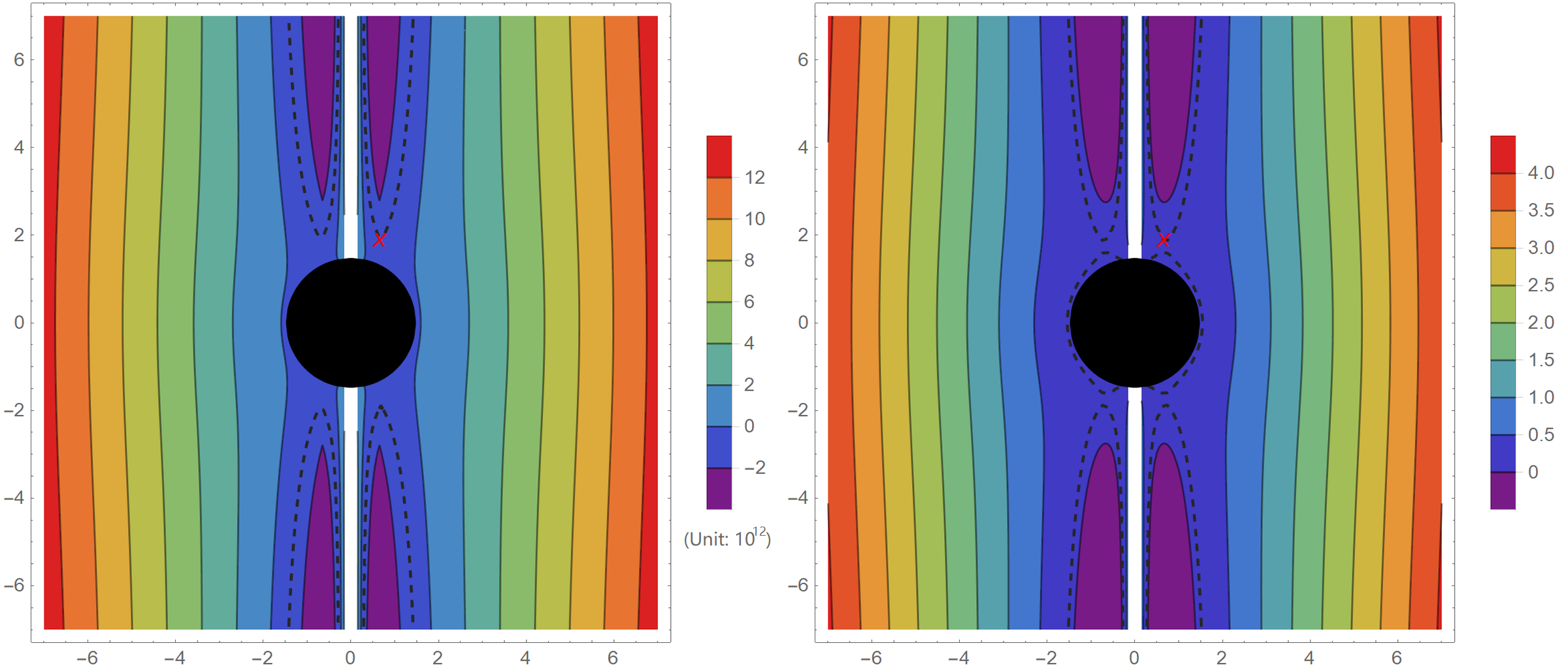}
  \centering
  \caption{The effective potential of charged particles along SVMs with $p_\phi(\tau=0)=0$. The horizontal and vertical coordinates are $\rho=r\sin\theta$ and $z=r\cos\theta$, respectively. The red cross marks the initial position of the charged particle in each plot. Left: $\omega_B=\omega_1=-5.59\times10^{12}$; Right: $\omega_B=\omega_2=-1.7$.}
  \label{jetVeff}
\end{figure}

\begin{figure}[h!]
  \centering
  \includegraphics[width=6in]{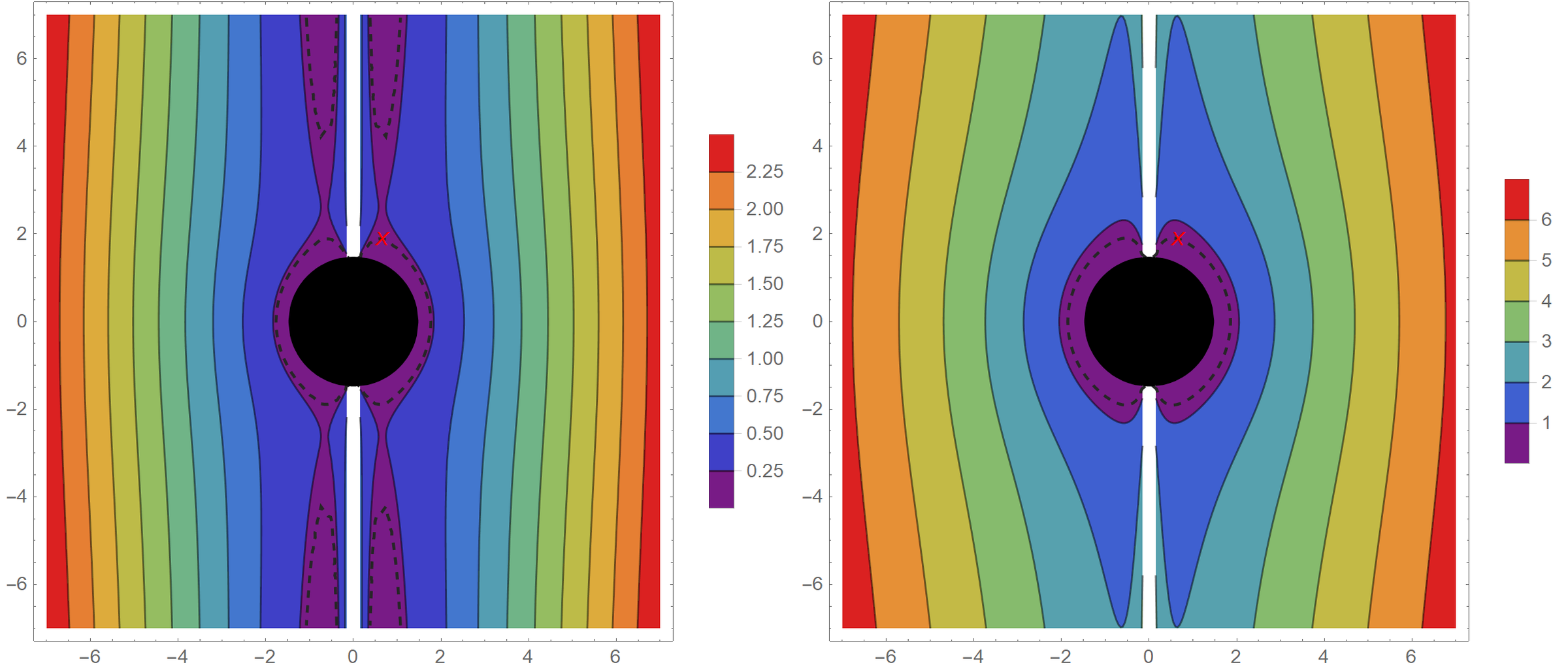}
  \centering
  \caption{The effective potential of falling charged particles with $p_\phi(\tau=0)=0$. The horizontal and vertical coordinates are $\rho=r\sin\theta$ and $z=r\cos\theta$. The red cross marks the initial position of the charged particle in each plot. Left: $\omega_B=-1$; Right: $\omega_B=+1.7$}
  \label{unjetVeff}
\end{figure}

Next, we would like to consider some specific situations. We set the spin of the Kerr BH to $a=0.94$, which is in the scope of the estimated values provided by the EHT Collaboration \cite{EventHorizonTelescope:2019pgp}. In this case, we set $r_i=2>r_+\simeq1.34$ and $\theta_i=\pi/9<\theta_c\simeq52.10^\circ$. Then we can solve $\partial_zV_\eff=0$ at $(r, \theta)=(2, \pi/9)$, and find $\omega_c\simeq-1.48$. Thus, for a charged particle released at $(r, \theta)=(2, \pi/9)$, we know that when $|\omega_B|>1.48$ the particle can be ejected from around the BH, while $|\omega_B|<1.48$ the charged particle would fall into the BH. In Fig. \ref{jetVeff}, we show the effective potentials $V_\eff$ at $\omega_B=\omega_1=-5.59\times10^{12}$ and $\omega_B=\omega_2=-1.7$. The red cross marks the initial position of the charged particle, and the grey dashed line is the contour of the effective potential where the charged particle lies at the initial time in each plot of Fig. \ref{jetVeff}. In addition, the values of $V_\eff$ are not shown in the white region since the values near the axis are too large. In the left plot of Fig. \ref{jetVeff}, we set $\omega_B=\omega_1$ and see that the effective potential above the initial position is lower and the motions of electrons are confined in the narrow stripe enclosed by the dashed line. Hence, electrons cannot be captured by the BH. Instead, they could spray out. Also, noting the axial symmetry of spacetime, the two stripes in the hemisphere are connected when considering the $\phi$-direction. In the right one of Fig. \ref{jetVeff}, we set $\omega_B=\omega_2$ and find that apart from the contours of the effective potential similar to the case at $\omega_B=\omega_1$ there are other contours of the effective potential around the BH. However, although the effective potentials inside the contours are all lower, the charged particles still cannot move toward the BH since there are barriers between the two lower potential areas. As a result, we can see that in both cases discussed above, charged particles have to go outward under the influence of the Lorentz force.

\begin{figure}[h!]
  \centering
  \includegraphics[width=5.5in]{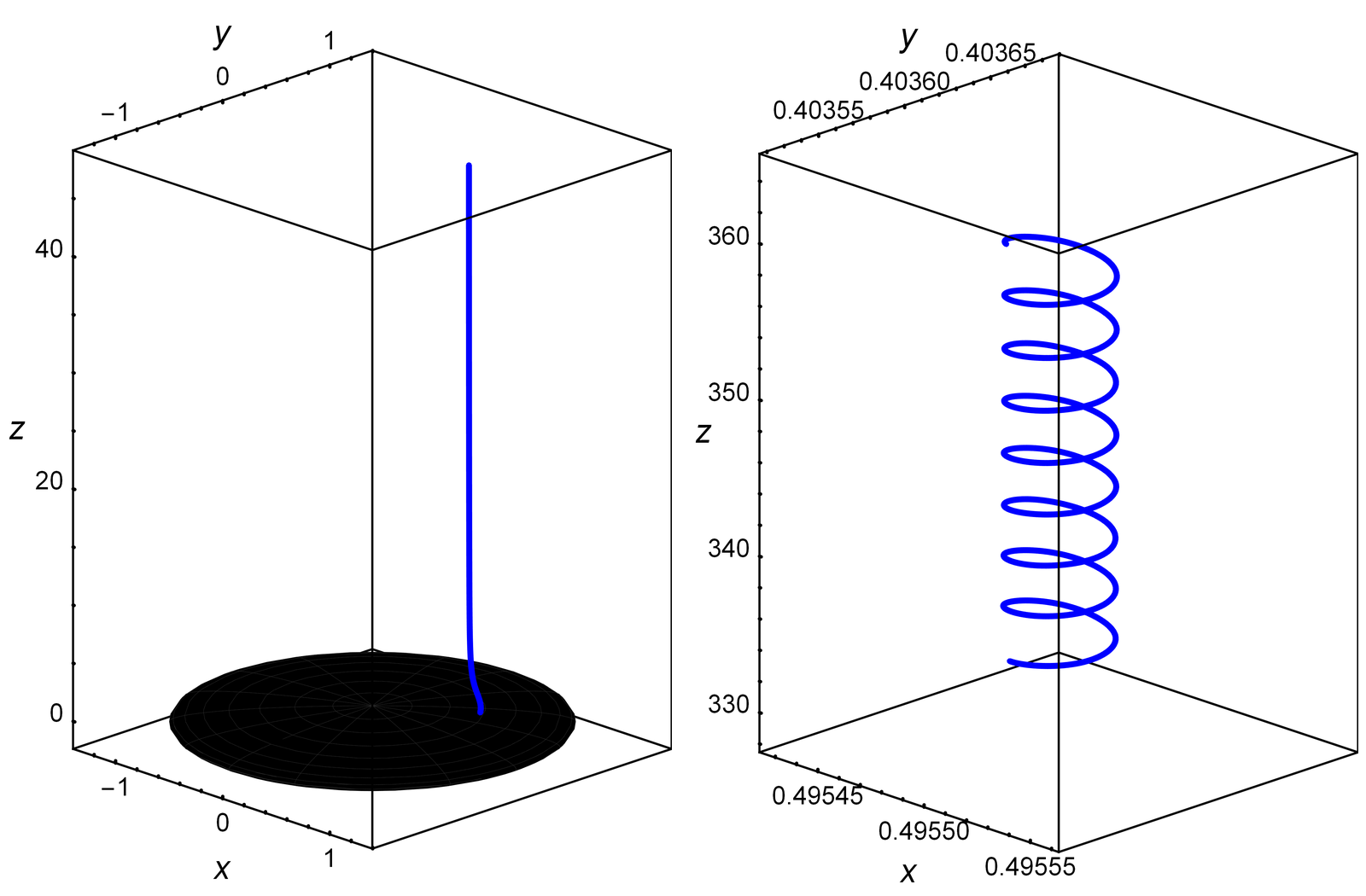}
  \centering
  \caption{Trajectory of charged particles along a SVM with $p_\phi(\tau=0)=0$ for $\omega_B=\omega_1$. Left: After a short curvilinear motion, the electron almost follows the direction of the magnetic field from a macro perspective. Right: Zoom in on the trajectory: it appears obviously spiraling upward when $z$ is large.}
  \label{wB559traj}
\end{figure}

\begin{figure}[h!]
  \centering
  \includegraphics[width=5.5in]{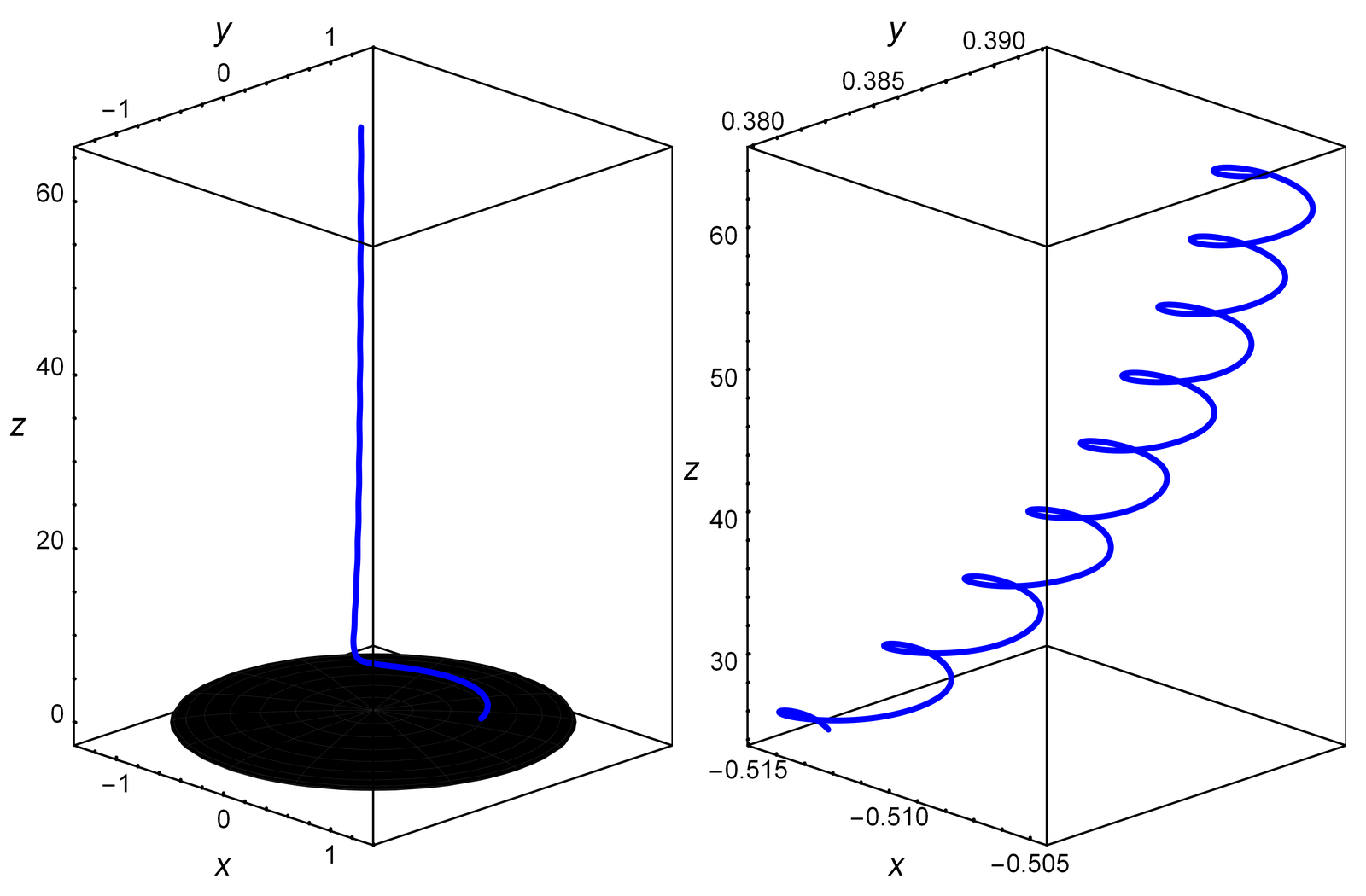}
  \centering
  \caption{Trajectory of charged particles along a SVM with $p_\phi(\tau=0)=0$ for $\omega_B=\omega_2$. Left: After a period of curvilinear motion, the electron also almost follows the direction of the magnetic field from a macro perspective. Right: Zoom in on the trajectory: it appears obviously spiraling and moving obliquely upward.}
  \label{wB17traj}
\end{figure}

\begin{figure}[h!]
  \centering
  \includegraphics[width=2.7in]{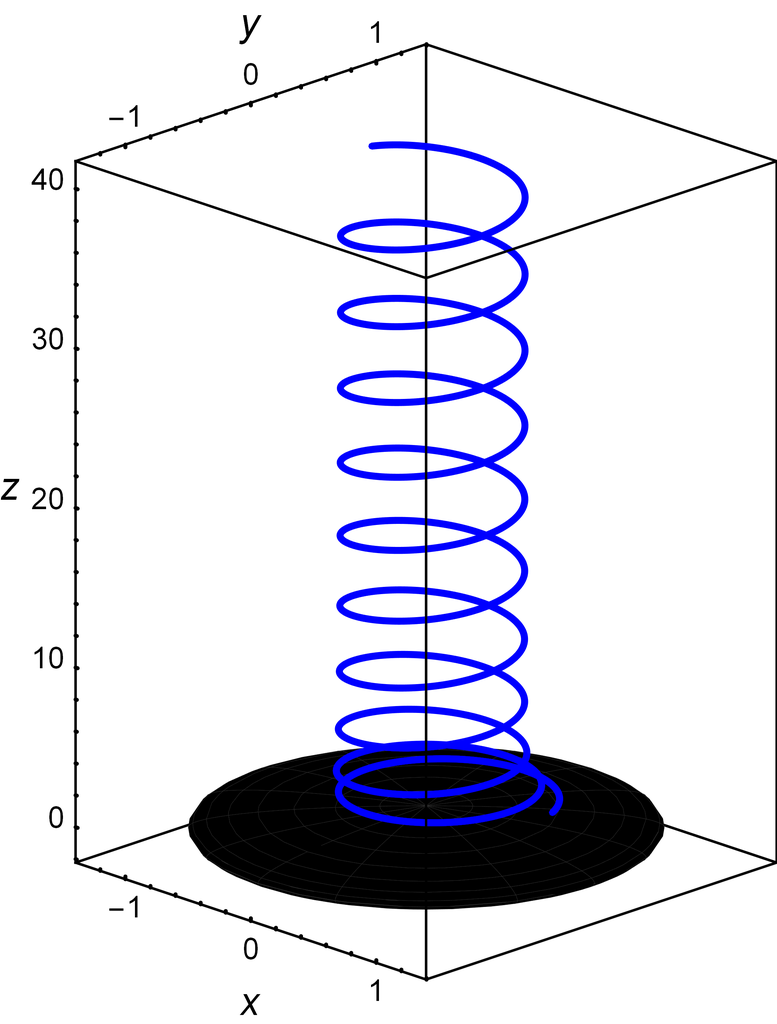}
  \centering
  \caption{Trajectory of charged particles along a SVM for $\omega_B=-3$. The charged particle is released with an initial momentum $p_\phi=1$. The magnitude of the spiral is comparable with the horizon of the BH.}
  \label{wB3traj}
\end{figure}
For comparison, we also show two examples in Fig. \ref{unjetVeff}, where we set $\omega_B=-1$ in the left plot and $\omega_B=+1.7$ in the right one. One can easily see that in the left plot although $\omega_B<0$, the charged particle has to fall into the BH since $|\omega_B|<1.48$ and the outward Lorentz force is not sufficient to resist the attraction of gravity. As for the right plot, obviously the charged particle cannot go ourward since $\omega_B>0$ and the Lorentz force is inwardly directed. 

In Fig. \ref{wB559traj} and Fig. \ref{wB17traj}, we present the trajectories of the motions of charged particles at $\omega_B=\omega_1$ and $\omega_B=\omega_2$ in the Kerr-Schild coordinates. Recall that the relation of spatial coordinates between the Kerr-Schild coordinates and the BL ones is 
\bea
x=\left(r\cos\phi-a\sin\phi\right)\,,\quad y=\left(r\sin\phi+a\cos\phi\right)\,,\quad z=r\cos\theta\,.
\eea 
Also, note that the charged particle is released at $(0, 2, \pi/9, 0)$ in BL coordinates with vanishing 3-velocity in the LNRF. We can see that each trajectory of charged particles for $\omega_B=\omega_1$ and $\omega_B=\omega_2$ starts with a curve and turns to a straight line at first glance. And the straight line is basically along the direction of the magnetic field. However, when we zoom in on the straight trajectory, we find it is actually spiral because of the magnetic field. The main reason why the degree of the spiral is small is that the velocity of the particle perpendicular to the direction of the magnetic field is small throughout the trajectory. In addition, as the specific charge $|q/m|$ increases, the Lorentz force on the charged particle increases, so that the spiral effect of the magnetic field will be more  suppressed. Comparing Fig. \ref{wB559traj} with Fig. \ref{wB17traj}, one can easily see that the spiral motion is more significant for $\omega_B=\omega_2$.
In order to see a trajectory of which the magnitude of the spiral is large enough to be comparable with the horizon of the BH, we set $\omega_B=-3$ and release the charged particle with an initial momentum $p_\phi=1$ at $(0, 2, \pi/9, 0)$ in the BL coordinates. The trajectory is shown in Fig. \ref{wB3traj}. One can see that the spiral motion upwards at the scale of the horizon of the BH and the rotation axis is very near the spin axis of the BH. Now we are convinced that the SVMs as expected can occur in the jet region under practical condition.

\section{Polarized images of SVM radiations}\label{sec3}
Considering that the accelerated charged particles in the magnetic field emit radiations synchronously, the SVMs in the jet region can provide polarized radiation profiles which can be observed by the EHT. Thus, in this section we move forward to study the polarized images of SVM radiations. A covariant method to evaluate the directions and intensities of electromagnetic radiations in a curved spacetime has been developed in our previous work \cite{Hu:2022sej}, which has been used to calculate the polarized images of synchrotron radiations emitted from circular motions in \cite{Zhu:2022amy, Lee:2022rtg} combining with the numerical backward ray-tracing method detailed in our another work \cite{Hu:2020usx}. In this work, we push our previous analysis in \cite{Hu:2022sej, Hu:2020usx} to obtain the polarized images of the SVM radiations. 

\subsection{Polarization, intensity and numerical method}

Now, we set up our problems and briefly introduce the numerical techniques that we need to use later. For simplicity, we assume the source comprises one or several trajectories. Each trajectory is formed by a charged ball-like object with a small constant radius $b$ in the BL coordinates moving along a SVM. We assume the radiations are emitted from the surface of the ball-like object, and the four-velocity of each point on the surface is approximatively described by the four-velocity of the center of the thing.
The images generated by our treatment would be similar to time-integrated images of a hotspot \cite{Broderick:2005my, Meyer:2006fd, Hamaus:2008yw, Dokuchaev:2020rye, GRAVITY:2020lpa, Rosa:2022toh, Rosa:2023qcv}.
In addition, in the limit $b\to 0$, the charged ball-like object reduces to a point-like charged particle. 

Following the results in \cite{Hu:2022sej}, an accelerated point-like charged particle in a magnetic field would produce electromagnetic radiations, whose polarization vector is given by 
\bea\label{fmu}
f^\mu=N^{-1} \left(K_\nu u^\nu a^\mu-K_\nu a^\nu u^\mu\right)\,,
\eea
where $K^\mu$ is the wave vector of the electromagnetic radiation, $N$ is a normalization factor, which is not important here since $f^\mu$ only encodes the directional information of the electromagnetic radiations. And $a^\mu$ is the $4$-acceleration of the charged particle, which takes 
\bea
a^\mu=\frac{q}{m}F^{\mu\nu}u_\nu=u^\nu\nabla_\nu u^\mu\,.
\eea
In addition, different from the model in \cite{Hu:2022sej} where the source is considered as a spot, in this paper the source is considered as the trajectories of charged body, which can be regarded as an extended light source. Thus, the expression of the emission intensity is slightly different from that appeared in \cite{Hu:2022sej}, which approximately takes
\bea\label{is}
I_s=\frac{4q^2}{b^2}\left[\left(\frac{K_\mu a^\mu}{K_\nu u^\nu}\right)^2+a^\mu a_\mu\right]\,.
\eea
Note that in Eqs. (\ref{fmu}) and (\ref{is}) we have reparameterized the wave vector $K^\alpha$ here compared to the wave vector $k^\alpha$ defined in \cite{Hu:2022sej}. More precisely, the relation between $K^\mu$ and $k^\mu$ reads 
\bea
k^\mu=-\frac{K^\mu}{K_\nu u^\nu}\,,
\eea
which follows the gauge $k^\mu u_\mu=-1$. Then, when we apply the numerical backward ray-tracing method to obtain the image of the electromagnetic radiations from the trajectories, we are able to set $K_\mu Z^\mu=1$ without loss of generality, where $Z^\mu$ is the $4$-velocity of the so-called zero-angular-momentum-observer (ZAMO) with the coordinates $(t_o, r_o, \theta_o, \phi_o)$. Here we want to stress that $K^\mu$ is past-directed vector in the backward ray-tracing procedure. The ``redshift'' factor between the source and the ZAMO is given by
\bea
g=\frac{K_\mu Z^\mu}{K_\mu u^\mu}=\frac{1}{K_\mu u^\nu}\,,
\eea
and then the intensity at the ZAMO is
\bea
I_o=g^4 I_s\,.
\eea
Note that we have integrated the observed frequency that give rise to $g^4$ in the above equation. Then, we normalize $I_o$ as 
\bea
T=\frac{1}{\log_2\left(1+\frac{1}{I_o/I_o^\ma}\right)}\,,
\eea
where $I_o^\ma$ is defined as the maximum value of $I_o$, so that $T$ falls between $0$ and $1$. It is worth stressing that when defining $T$ we refer to the expression of the brightness temperature \cite{rybicki1991radiative}, which is defined as the measured temperature of the emissions from a black body. To image the source on the screen of the ZAMO, we employ the the stereographic projection by introducing standard Cartesian coordinates $(\X, \Y)$ on the screen, which is also called the fisheye camera model. To begin with, we define the frame of the ZAMO, whose tetrad $\{\hat{e}^\nu_{(\mu)}\}$ 
can be found from Eqs. (3.5) to (3.8) in \cite{Zhong:2021mty}
\bea
\hat{e}_{(0)}=\frac{g_{\phi\phi}\partial_t-g_{\phi t}\partial_{\phi}}{\sqrt{g_{\phi\phi}\left(g_{\phi t}^2-g_{\phi\phi}g_{tt}\right)}}\,,\quad\quad \hat{e}_{(1)}=-\frac{\partial_r}{\sqrt{g_{rr}}}\,,\quad
\hat{e}_{(2)}=\frac{\partial_\theta}{\sqrt{g_{\theta\theta}}}\,,\quad \hat{e}_{(3)}=-\frac{\partial_\phi}{\sqrt{g_{\phi\phi}}}\,,
\eea
which are very similar to the tetrad of the LNRF in Eq. (\ref{tetrad}). Note that the minus signs in $\hat{e}_{(1)}$ and $\hat{e}_{(3)}$ are added to facilitate the backward ray-tracing method, that is, $\hat{e}_{(1)}=-e_{(1)}$ and $\hat{e}_{(3)}=-e_{(3)}$. The details of the backward ray-tracing method used in this work can be found in the Appendix. B of \cite{Hu:2020usx}. Following the convention in \cite{Hu:2020usx}, the celestial coordinates are defined as
\bea
\cos\Theta=\frac{K_\mu \hat{e}^\mu_{(1)}}{|K_\mu \hat{e}^\mu_{(0)}|}\,,\quad \tan\Psi=\frac{K_\mu\hat{e}^\mu_{(3)}}{K_\mu\hat{e}^\mu_{(2)}}\,,
\eea
and the relation between the celestial coordinates and the standard Cartesian coordinates $(\X, \Y)$ is given by \cite{Hou:2022eev}
\bea
\X=-2\tan\frac{\Theta}{2}\sin\Psi\,,\quad \Y=-2\tan\frac{\Theta}{2}\cos\Psi\,.
\eea
In addition, the information of $f^\mu$ given at the source can be decoded on the screen of the ZAMO with the aid of the Penrose-Walker (PW) constant $\kappa$ \cite{Lupsasca:2018tpp} in the Kerr spacetime, which takes
\bea\label{dip}
\vec{\mathcal{E}}=\left(\E_\X, \E_\Y\right)=\frac{1}{K_t\left(\Y_o^2+\Z_o^2\right)}\left(\kappa_2\Y_o-\kappa_1\Z_o, \kappa_1\Y_o+\kappa_2\Z_o\right)\,,\quad \Z_o=-\left(\X_o+a\sin\theta_o\right)\,,
\eea
and 
\bea\label{kappa}
\kappa_1+i\kappa_2\equiv \kappa=-2K^\mu f^{\nu}\left(\hat{l}_{[\mu}\hat{n}_{\nu]}-\hat{m}_{[\mu}\hat{\bar{m}}_{\nu]}\right)(r-ia\cos\theta)\,,
\eea
where $\{\hat{l}, \hat{n}, \hat{m}, \hat{\bar{m}}\}$ are the Newman-Penrose tetrads whose explicit expressions can be found from Eq. (10) in \cite{Lupsasca:2018tpp}. Obviously, Eq. (\ref{kappa}) can be calculated at the source and then one can obtain the direction of the polarization on the screen from Eq. (\ref{dip}) considering that $\kappa$ is a constant along a null geodesic in Kerr spacetime.

\begin{figure}[h!]
  \centering
  \includegraphics[width=7in]{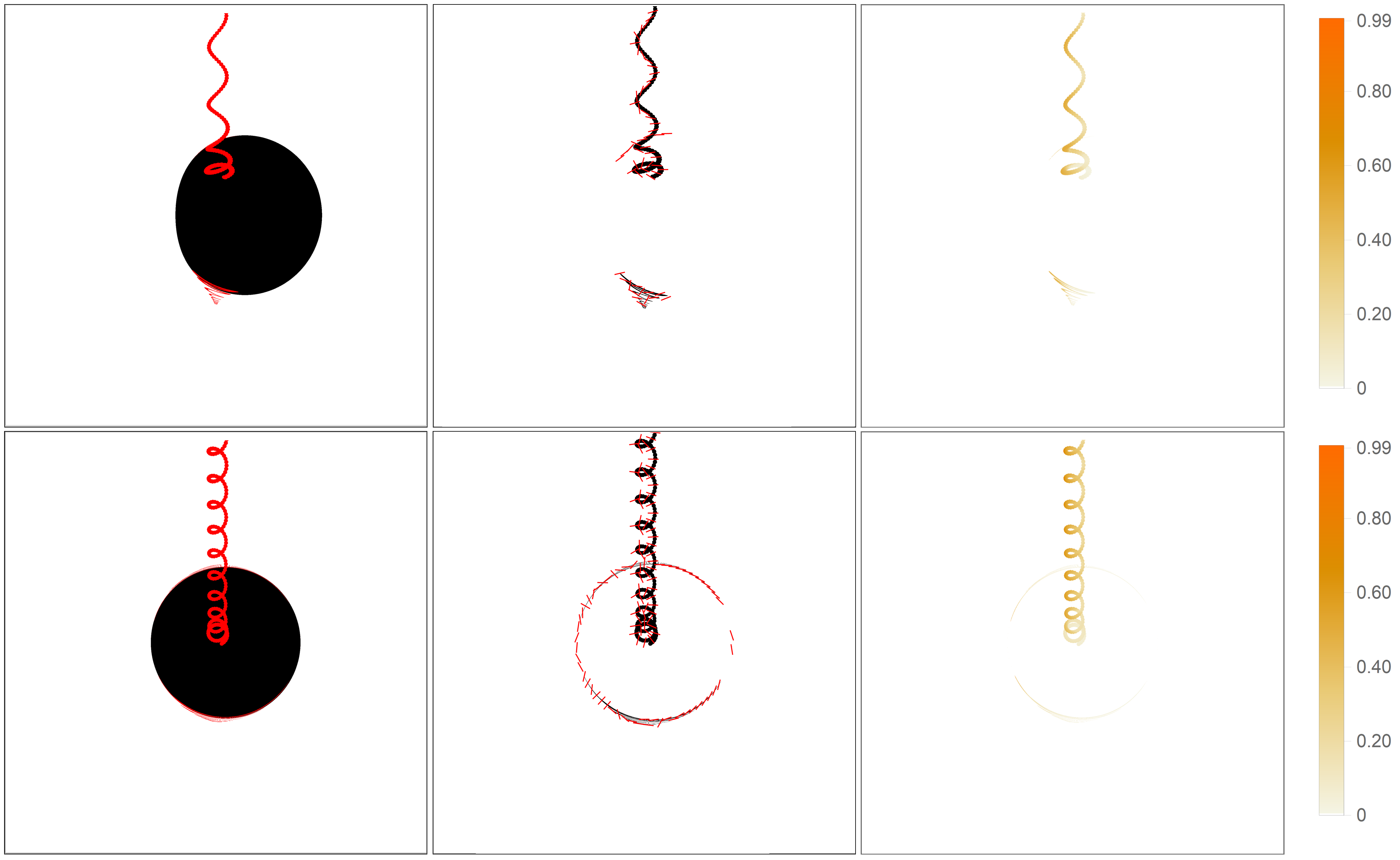}
  \centering
  \caption{Images of a single trajectory along the SVM with $\omega_B=-3$. Charged particles are released at $r_s=2, \theta_s=\pi/9, \phi_s=0$ with an initial momentum $p_\phi=1$ in the northern hemisphere. Left: Images with the BH shadow; Middle: Polarization directions; Right: The intensity of the image; Upper: $\theta_o=90^\circ$; Lower: $\theta_o=17^\circ$.}
  \label{sB}
\end{figure}

\subsection{Results}

Now, we are ready to explore the images of the charged source in the jet region. Since the BH is very far from us in the universe, we set $r_o=300\gg r_h$ in our numerical program. In terms of the Killing vectors $\partial_t$ and $\partial_\phi$ in the spacetime, we fix $t_o=\phi_o=0$ in the following. As for the observational angle, we focus on two situations: one is that the observer is located on the equatorial plane, that is $\theta_o=90^\circ$, and the other one is that we set $\theta_o=17^\circ$ which corresponds to the observational angle of viewing the supermassive BH in the center of M87 galaxy. Recall that the spin of the Kerr BH and the magnitude of the magnetic field have been chosen as $a=0.94$ and $B_0=10 ~\text{Gauss}$ in studying the SVMs in the subsection \ref{2sub2}, we set $b=0.15\ll r_h\simeq1.34$. The remaining parameter that is not determined is the specific charge or say the parameter $\omega_B=qB_0/m$, which was defined in the subsection \ref{2sub2}. In this work, our main interest is to see the images of the trajectories along the SVMs of electrons,  we will have an investigation on the images for $\omega_B=\omega_1$ with a vanishing initial velocity later. In addition, considering that the SVM becomes visible at the scale of the horizon of the BH when $\omega_B=-3$ with $p_r(\tau=0)=p_\theta(\tau=0), p_\phi(\tau=0)=1$, we would like to first make some explorations in this simple case.

\begin{figure}[h!]
  \centering
  \includegraphics[width=7in]{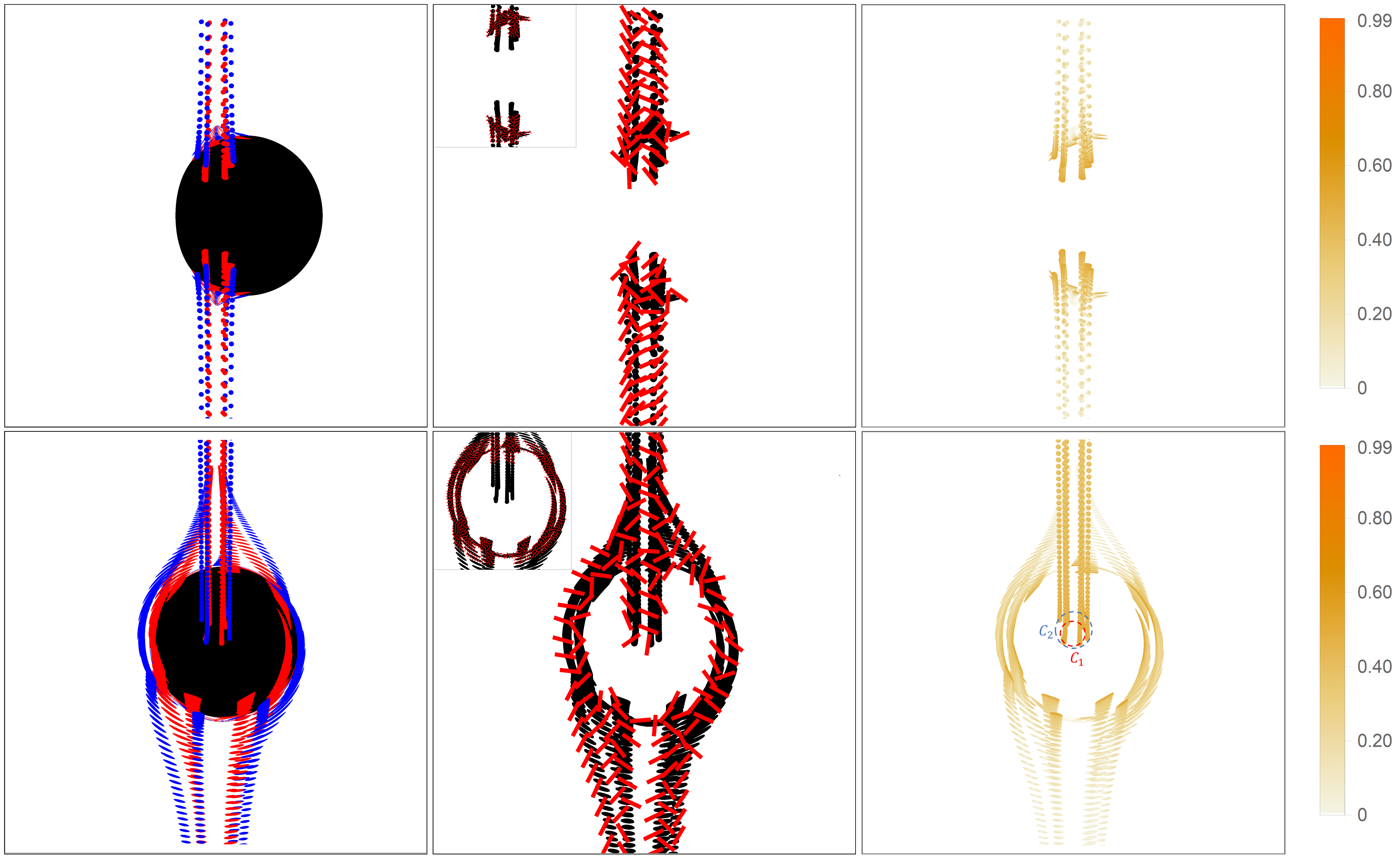}
  \centering
  \caption{Images of sixteen trajectories along the SVM with $\omega_B=\omega_1$. The sixteen trajectories are symmetric pairwise about the equatorial plane. In the northern hemisphere, four charged particles of the trajectories in red are released at $r_s=2, \theta_s=\pi/9, \phi_s=0,\,\pi/2,\,\pi,\,3\pi/2$ and the other fours of the trajectories in blue are released at $r_s=3, \theta_s=\pi/9, \phi_s=\pi/4,\,3\pi/4,\,5\pi/4,\,7\pi/4$. Left: Images with the BH shadow; Middle: Polarization directions; Right: Intensity of the images; Upper: $\theta_o=90^\circ$; Lower: $\theta_o=17^\circ$. We zoom in near the critical curve in the polarized images.}
  \label{r2r3}
\end{figure}

In Fig. \ref{sB}, we show the images for the case of the apparent spiral motion observed at $\theta_o=90^\circ$ and $\theta_o=17^\circ$. More precisely, we have $\theta_o=90^\circ$ for the upper panel of Fig. \ref{sB} and $\theta_o=17^\circ$ for the lower one. On the left panel, we show the BH shadows corresponding to null geodesics falling into the horizon and shapes of the images of the source. On the middle board, we offer the polarization directions of the electromagnetic radiations along the trajectories. On the right panel, we show the intensities of images. The source is a single trajectory along the SVM presented in the Fig. \ref{wB3traj}, where the charged particles with $\omega_B=-3$ are released at $r_s=2, \theta_s=\pi/9, \phi_s=0$ with a nonvanishing initial momentum $p_\phi=1$. From Fig. \ref{sB}, we can see the primary image of the source in the northern hemisphere of each plot both for $\theta_o=90^\circ$ and $\theta_o=17^\circ$. A demagnified secondary image can be seen clearly in the southern hemisphere, and higher order images are barely visible around the shadow curve at $\theta_o=90^\circ$. While at $\theta_o=17^\circ$, higher order images are visual, although the secondary image is less clear. As a result, we can roughly observe a photon ring structure in the photo at $\theta_o=17^\circ$, which implies that the photon rings might be universally observed when the SVM radiation serves as the light source, without involving the accretion disc. In addition, we can see a big difference between the shapes of the primary images observed at $\theta_o=90^\circ$ and $\theta_o=17^\circ$ due to the strong gravitational lensing around the BH. Furthermore, we can see that the intensity on the left side of the trajectory is more significant than that on the other side because the object moves in a counterclockwise spiral in the view of the north pole. When the source reaches the left side, its velocity flows toward the observer. On the contrary, when it arrives at the right side, its velocity is opposite to the observer's. Due to the Doppler's effect, the intensity on the left side is larger than the one on the other side. Moreover, the polarization directions along the trajectories change periodically for the primary images.

In Fig. \ref{r2r3} we show the images for the radiations from the electrons in SVMs observed at $\theta_o=90^\circ$ and $\theta_o=17^\circ$. In this case we have $\omega_B=\omega_1$. Recall that the motion trail of a SVM electron is almost a straight line and unchanged along $x$ and $y$ directions on a macro scale as presented in Fig. \ref{wB559traj}, in order to have a better imaging result we set the light source to multiple trajectories along the corresponding SVMs in this situation. In the northern and southern hemispheres, we place sixteen trajectories pairwise-symmetrically as the light source. And the eight trajectories in the northern hemisphere start at a conical surface with the half-angle of projection $\theta_s=\pi/9$. Furthermore, we set half of the starting points at $r_s=2$ with $\phi_s=0,\,\pi/2,\,\pi,\,3\pi/2$, and the other half at $r_s=3$ with $\phi_s=\pi/4,\,3\pi/4,\,5\pi/4,\,7\pi/4$, so that their images can be distinguished as much as possible on the screen of the observers. Then considering the $Z_2$ symmetry of the Kerr BH spacetime, the trajectories in the southern can be understood easily. Similar to Fig. \ref{sB}, we also show the images observed at $\theta_o=90^\circ$ on the upper panel and the lower panel gives the results for $\theta_o=17^\circ$. The left two plots illustrate the BH shadows and the images of trajectories, where the red colour signifies the results for $r_s=2$ and the blue one denotes the results for $r_s=3$. In the middle plots we show the polarization directions along the images, and in the right plots we present the intensities of the images. We zoom in near the critical curve in the polarized images and show the detailed images in the upper-left corner of the middle plots.

According to Fig. \ref{r2r3}, the images obtained by the equatorial observers remain $Z_2$ symmetric, and the intensity of primary images decreases as the particles move farther away from the BH. Furthermore, the high-order images for equatorial observers cannot form a complete ring structure because of the limitation of $\theta$-oscillation of null geodesics \cite{Gralla:2019drh}. In contrast, a photon ring structure can be formed by the primary and high-order images of the SVM source in the screen of an observer at $\theta_o=17^\circ$. Due to the Doppler effect, the radiations from the northern hemisphere are much brighter than those in the southern hemisphere. In addition, we draw two circles $C_{1,2}$ in the lower right plot in Fig. \ref{r2r3}, which cross the starting points of four trajectories for $r_s=2$ and $r_s=3$, respectively. One can infer that all the starting points of trajectories for $r_s=2$ and $\phi_s\in[0, 2\pi]$ could form a closed curve close to $C_1$ and something similar is true for $C_2$. Thus we can imagine that the trajectories starting at a larger value of $r_s$ lead to a larger circle, and they may occupy the whole BH shadow region if charged particles are ejected at each $r_s\in(r_h, \infty)$. As a result, we may lose the so-called BH shadow. The main reason for the disappearance of BH shadow is that our source is assumed to be opaque. One can expect a significant intensity gap if the SVM source becomes optically thin. Moreover, for the observers at $\theta_o=90^\circ$ and $\theta_o=17^\circ$, the polarization of high-order images is indistinct and in disorder while the primary images have specific approximately parallel polarization directions.

\section{Summary and discussion} \label{sec4}

In this present paper, we studied the polarized images of SVM radiations near the Kerr BH with a vertical and uniform magnetic field, which is described by the Wald solution. Our work mainly includes two aspects. In the first part, we focused on the SVMs of charged particles and found a critical value $\omega_c$ of the parameter $\omega_B=qB_0/m$ when charged particles were subjected to an outward electric field force in the LNRF. In the jet region, that is $0<\theta<\theta_c$, the charged particles at rest in the LNRF would be ejected when $|\omega_B|>|\omega_c|$, or they would fall into the BH. Then, we presented the trajectories of SVMs for charged partilces with various $\omega_B$ including the electrons. 

In the second part, we investigated the polarized images of the radiations from the trajectories along the SVM at $\theta_o=90^\circ$ and $\theta_o=17^\circ$. As a warm up, we showed the image of a single trajectories along the SVM with $\omega_B=-3$, which were released at $r_s=2, \theta_s=\pi/9, \phi_s=0$ with an initial momentum $p_\phi=1$. Then, we considered sixteen trajectories along the SVM charged particles with $\omega_B=\omega_1$, which were symmetric about the equatorial plane. From the images observed at $\theta_o=90^\circ$ and $\theta_o=17^\circ$, we found that the SVM radiations could form the photon ring structure  without the radiations from the accretion disk.

In our model, the images are quite different from the pictures of the supermassive BHs in M87 and Milky Way galaxies except for the photon ring structure. There are two main reasons, as follows. First, the covariant method we used to calculate intensity in \cite{Hu:2022sej} is independent of photon frequency, while a real observation must include the effects of the radiation frequency. Hence, the contributions of SVM radiations may be not significant when observed at $230$ GHz, which is applied in the observation of EHT. Secondly, the electromagnetic radiation of charged particles is likely weaker than the thermal radiations of the electrons in the fluid mode; that is, the signals of SVM radiations may be overwhelmed by the thermal radiations. 

Nevertheless, our results are still of potential astrophysical interest. It is known that most of the synchrotron power is radiated around the peak of the spectrum which occurs near the characteristic frequency $\nu_c \sim qB_0\gamma^2/m$, where $\gamma$ is the Lorentz factor of the emitter, measured by the observer at rest at infinity. For electron, under extremly strong magnetic field, most of the synchrotron power is in the gamma-ray band since the charge to mass ratio $e/m$ is large \cite{Rueda:2022fgz}. However, for supermassive black holes such as M87$^\ast$ and Sgr A$^\ast$ with relatively mild magnetic fields, the characteristic frequency of synchrotron radiation from ejected electrons may be significantly suppressed. In addition, a charged object with smaller charge to mass ratio, such as the charged dust, will also lead to a lower characteristic frequency. Thus, we can expect to observe the radiation signal of SVMs from the EHT or near-infrared flares \cite{abuter2018detection} in the future.

\section*{Acknowledgments}
We thank Jiewei Huang, Changkai Luo and Ye Shen for helpful discussions. The work is partly supported by NSFC Grant No. 12275004, 12205013 and 11873044. MG is also endorsed by ”the Fundamental Research Funds for the Central Universities” with Grant No. 2021NTST13.

\appendix

%

\end{document}